\documentclass[conference]{IEEEtran}
\IEEEoverridecommandlockouts
\usepackage{cite}
\usepackage[numbers,sort&compress]{natbib}

\usepackage[utf8]{inputenc}
\usepackage{amsmath,amssymb}
\usepackage{graphicx}
\usepackage{bm}
\usepackage{enumerate}
\usepackage{comment}
\usepackage{xcolor}
\usepackage{url}
\usepackage{color,soul}
\usepackage{subcaption}
\usepackage{float}

\definecolor{light-gray}{gray}{0.8}

\usepackage{amsmath,amssymb,amsfonts}
\usepackage{algorithmic}
\usepackage{graphicx}
\usepackage{textcomp}
\usepackage{xcolor}
\usepackage{subcaption}

\def\BibTeX{{\rm B\kern-.05em{\sc i\kern-.025em b}\kern-.08em
    T\kern-.1667em\lower.7ex\hbox{E}\kern-.125emX}}

\makeatletter
\newcommand{\linebreakand}{%
  \end{@IEEEauthorhalign}
  \hfill\mbox{}\par
  \mbox{}\hfill\begin{@IEEEauthorhalign}
}
\makeatother

\begin{document}

\title{Research on Cloud Platform Network Traffic Monitoring and Anomaly Detection System based on Large Language Models\\}

\author{

\small 

\begin{tabular}[t]{c@{\extracolsep{8em}}c} 

1\textsuperscript{st} Ze Yang \textsuperscript{*}  & 2\textsuperscript{nd} Yihong Jin \\
\textit{University of Illinois Urbana-Champaign} & \textit{University of Illinois Urbana-Champaign} \\
Champaign, IL 61801, USA & Champaign, IL 61801, USA \\
\textsuperscript{*}Corresponding Author: zeyang2@illinois.edu & yihongj3@illinois.edu\\

\\

3\textsuperscript{rd} Juntian Liu & 4\textsuperscript{th} Xinhe Xu \\
\textit{Computer Science Department} & \textit{Computer Science Department} \\
\textit{University of Illinois Urbana-Champaign} & \textit{University of Illinois Urbana-Champaign} \\
Champaign, IL 61801, USA & Champaign, IL 61801, USA \\
jl203@illinois.edu & xinhexu2@illinois.edu  \\

\\

5\textsuperscript{rd} Yihan Zhang & 6\textsuperscript{th} Shuyang Ji \\
\textit{Computer Science Department} & \textit{Computer Science Department} \\
\textit{University of Illinois Urbana-Champaign} & \textit{University of Illinois Urbana-Champaign} \\
Champaign, IL 61801, USA & Champaign, IL 61801, USA \\
yihanz8@illinois.edu & sji15@illinois.edu  \\

\\
\end{tabular}
}

\maketitle

\begin{abstract}
The rapidly evolving cloud platforms and the escalating complexity of network traffic demand proper network traffic monitoring and anomaly detection to ensure network security and performance. This paper introduces a large language model (LLM)-based network traffic monitoring and anomaly detection system. In addition to existing models such as autoencoders and decision trees, we harness the power of large language models for processing sequence data from network traffic, which allows us a better capture of underlying complex patterns, as well as slight fluctuations in the dataset. We show for a given detection task, the need for a hybrid model that incorporates the attention mechanism of the transformer architecture into a supervised learning framework in order to achieve better accuracy. A pre-trained large language model analyzes and predicts the probable network traffic, and an anomaly detection layer that considers temporality and context is added. Moreover, we present a novel transfer learning-based methodology to enhance the model's effectiveness to quickly adapt to unknown network structures and adversarial conditions without requiring extensive labeled datasets. Actual results show that the designed model outperforms traditional methods in detection accuracy and computational efficiency, effectively identify various network anomalies such as zero-day attacks and traffic congestion pattern, and significantly reduce the false positive rate. 
\end{abstract}


\section{Introduction}
The cloud computing platform can combine flexible resource control, pay-as-you-go services, and efficient computing capabilities to become the core infrastructure for the transformation of enterprise information. Because the cloud platform can efficiently process massive data and support multiple data storage and analysis methods, it has become the core of modern enterprise digital operations \cite{olateju2024combating}. At the same time, though, cloud platforms are under new and unprecedented security pressures. 

As cloud computing environments are increasingly open and shareable, hackers and malicious attackers have become more sophisticated, and their attack targets have gradually shifted from the traditional LANs to cloud platforms. So how to take timely measures to monitor and analyze the network traffic in the cloud platform and find potential security risks timely becomes a central work to ensure that the platform operates stably and data is safe \cite{al2023intrusion}. The detection of abnormal network activity not only starts from monitoring network traffic to alert common attacks on cloud platforms by singing security threats such as the spread of computer viruses and DDoS attacks but also handles more complex and stealthy attack models.

Traditional network traffic monitoring methods are based on rule-based detection systems, which set predefined rules and thresholds and determine whether network traffic is anomalous. For instance, alerting when a specific volume of data traffic exceeds a known threshold \cite{alghamdi2021deep}. There are some known threats where this type of approach is effective, but we often find it very difficult to adapt to the rules based approach when the characteristics of network traffic change or new attack vectors emerge. 

Meanwhile, due to the diversification of service types and different user needs on the cloud platform, the form of network traffic has become more complex, and the traditional methods have presented very great limitations in dynamic and high-dimensional traffic data monitoring. Especially under complex variant attack scenarios (such as encrypted traffic and DDoS distributed attacks), rules and threshold methods usually cannot make effective accurate detection of these abnormal behaviors, leading to security protection blind areas \cite{lin2024enhanced}.

In the past years, deep learning and machine learning technologies have been broadly applied to the field of cyber security, especially to the anomaly detection methods based on sizable data and complex patterns. Machine learning and deep learning techniques have rapidly evolved and ushered unprecedented changes to the cybersecurity landscape. Machine learning does this better than traditional rule-based detection as it is able to find complex patterns buried within massive volumes of data through learning from large amounts of historical data. 
Due to their efficiency in feature extraction from the time series data and classifying it, deep learning, particularly CNNs and RNNs, have been extensively applied in network traffic analysis, deception detection, and intrusion detection \cite{antonini2023adaptable, li2024deception}. Not only these methods can significantly get the increase of detection accuracy, furthermore, they can also greatly reduce the false positive rate, in the high-dimensional and non-linear relationship data generally face, machine learning and deep learning methods have shown a powerful advantage. 
Therefore, how to use these advanced technologies to detect the abnormal behavior of network traffic has become the focus of recent research, especially for complex network environment and dynamic attack behaviors, which has achieved certain results. But theoretically speaking on the one hand although machine learning methods are very likely to have that effect but on the other hand is how to really solve the problem of real-time performance, scalability, and unforeseen attacks, these have become an urgent need to be addressed issues.

\section{RELATED WORK}
Above all, Xu et al. \cite{xu2024towards} introduced SpeCoder, a framework that leverages speculative sampling to accelerate server-side code generation. This work highlights the growing importance of addressing the computational demands of LLMs and ensuring their responsiveness for time-sensitive tasks\cite{ni2024harnessing, ni2024timeseries}.

Yu et al. \cite{yu2024deep} proposes a deep-learning based anti-money laundering system to detect anomalies attached to cross-border transactions. Motivated by the fact that contrastive learning technology is empowering unsupervised learning models to significantly enhance the performance of cross-border AML systems and/or to optimize detection rules, this study formulates a hypothesized proposition studied through targeted experiments. Vervaet \cite{vervaet2021monilog} introduces Monilog, a log-based anomaly detection system that runs automatically for cloud computing infrastructure. By seamlessly integrating cutting-edge machine learning technologies, Monilog is capable of analyzing and processing large volumes of log data generated by cloud environments to detect behaviors. Ji et al. \cite{ji-etal-2024-rag} present a framework that uses Retrieval-Augmented Generation (RAG) and readability control to summarize complex biomedical texts for a general audience, showcasing advanced NLP techniques for information processing, which is conceptually related to how LLMs can process complex network traffic data.

Additionally, Alshammari and Aldribi \cite{alshammari2021apply} Detect malicious network traffic for cloud distributed environment using deep neural networks: A review. Thapa and Arjunan \cite{thapa2024ai} proposed the improved AI cybersecurity solution that uses machine learning to leverage anomaly detection in cloud environments. Torabi et al. \cite{torabi2023practical} focus on autoencoder-based anomaly detection method which uses vector reconstruction errors to detect anomaly in the cloud computing networks. The literature indicates that the recent advent of specific types of technologies including deep learning~\cite{preprec, li2024exploring, meng2024exchangerate}, unsupervised learning~\cite{infomotif, protocf}, autoencoders, and hybrid architectures for efficient detection in complex domains \cite{lu2022cot} has greatly enhanced the efficiency and accuracy of such anomaly detection methods in cloud environments.

\section{METHODOLOGIES}
\subsection{Network traffic data modeling}

In the network traffic monitoring and anomaly detection tasks, high-dimensional modeling of network traffic data is required. Suppose that at time $t$, the eigenvector of network traffic is represented as $X_t \in \mathbb{R}^n$, where $n$ is the dimension of the feature, and $X_t = \begin{bmatrix} x_{t1} & x_{t2} & \dots & x_{tn} \end{bmatrix}^T$ are the characteristics of all network traffic at that time $t$. Each characteristic $x_{ti}$ represents a key metric in network traffic (e.g., bandwidth usage, latency, packet loss rate, etc.). These features may exhibit complex temporal correlations at different time points $t$, so they need to be constructed. In order to capture these complex temporal dependencies, we first normalize the network traffic data $X_t$ to ensure that the dimensions of each feature are consistent, so that the model will not be affected by features of different magnitudes during training, as shown in Equation 1:

\begin{equation}
X_t' = \frac{X_t - \mu_t}{\sigma_t}
\label{eq:1}
\end{equation} where $\mu_t$ and $\sigma_t$ are the mean and standard deviation of each feature at time point $t$, respectively. The normalized data $X_t'$ is used as input to the subsequent model.

Subsequently, we design a new time-series dependency capture mechanism, which uses a self-attention mechanism to deal with the time series relationship in network traffic data. Suppose network traffic is in a time series; the feature matrix is $X \in \mathbb{R}^{T \times n}$, and our goal is to learn the interrelationships between time points through the attention mechanism. We use the self-attention mechanism in the Transformer architecture to calculate the degree of influence between time point $t$ and other time point $t'$. Specifically, the attention matrix $A \in \mathbb{R}^{T \times T}$ is defined as the relationship between time points in the network traffic data series, where each element $a_{tt'}$ represents the similarity between time point $t$ and $t'$, denoted as Equation 2:

\begin{equation}
A_{tt'} = \text{Softmax} \left( \frac{(X_t W_q) (X_{t'} W_k)^T}{\sqrt{d_k}} \right)
\label{eq:2}
\end{equation} where $W_q$ and $W_k$ are query matrix and key matrix, respectively, and $d_k$ is the scaling factor. Through the self-attention mechanism, the model can automatically learn the dependencies between time points. Finally, the model uses a weighted sum to obtain the representation $Z_t$ for each time point is Equation 3:

\begin{equation}
Z_t = \sum_{t'} A_{tt'} X_{t'}
\label{eq:3}
\end{equation}

This process can effectively capture the complex dynamic changes of network traffic in the time dimension, and ensure that the global information of historical time series can be taken into account when anomaly detection.

To address the issue of the model dominating an effective dynamic adaptation mechanism under rapid network structure and attack pattern changes, the addition of meta-learning allows the model to update parameters continually, using only new network traffic data, without performing full retraining, making it capable of recovering from sudden new attacks as well as large modifications in network topology. This structural adaptability is also reflected in recent work on transferable graph autoencoders, such as T-GAE \cite{tgae}, which shows strong potential in aligning evolving network representations across domains. Similar to methods used in biological data analysis under uncertain or missing signals \cite{du2024embracing}.

\subsection{Anomaly detection}
After capturing the time-series dependencies of network traffic, we integrate anomaly detection tasks into the overall system, and in order to accurately detect anomalous behavior in network traffic, traditional anomaly detection methods are often based on reconstruction errors, but this approach may ignore latent patterns in the data. In order to enhance the anomaly detection ability of the model, we propose an anomaly detection loss function based on temporal series self-attention by combining the auto encoder and supervised learning method. Assuming that the time series features obtained by the self-attention mechanism are represented as $Z_t$ and the reconstruction result of network traffic is $\hat{Z}_t$, the number of losses for anomaly detection is Equation 4:

\begin{equation}
L_{\text{detect}} = \sum_{t=1}^{T} \left( \|Z_t - \hat{Z}_t\|_2^2 + \lambda_t \|A_t\|_1 \right),
\label{eq:4}
\end{equation}

where $\|Z_t - \hat{Z}_t\|_2^2$ is the reconstruction error of the traffic data, which measures the ability of the model to restore the normal traffic pattern, $\lambda_t$ is the regularization factor, which controls the weight of the anomaly detection items, and $A_t$ is the time series relationship matrix calculated based on the attention mechanism. The loss function jointly optimizes the reconstruction error and anomaly detection terms of network traffic, so as to distinguish between normal and abnormal traffic.

We also adopted a transfer learning approach, which allows the model to quickly adapt to new network environments and attack patterns. With transfer learning, we migrate models that have been trained on source tasks to target tasks, avoiding the high cost of training from scratch. Specifically, we use the Multi-Task Learning framework for migration, assuming that the loss function of the source task is $L_{\text{source}}$ and loss function of target task is $L_{\text{target}}$, the overall loss function is Equation 5:

\begin{equation}
L_{\text{transfer}} = L_{\text{source}} + \sum_{i=1}^{m} \lambda_i L_{\text{target},i},
\label{eq:5}
\end{equation} where $m$ is the number of target tasks, and $\lambda_i$ is the regularization factor for each target task. With transfer learning, the model is better able to adapt to new network environments and attack scenarios without the need for large amounts of annotated data. 

Then with contrastive learning applied, the feature discrimination ability will improve to reliably identify such stealthy attacks. Concurrent use of Bayesian optimization to reduce the complexity of manual adjustment of parameters and automatically adjust the weights of the regularization factor and the time series relationship matrix in the loss function.

\section{EXPERIMENTS}
\subsection{Experimental setup}

We adopted the CICIDS 2017 dataset offered by Canadian Cyber Security Institute, which includes multiple cyber attacks including DDoS attacks, SQL injection, brute force attacks, etc. and normal traffic, and has been applied in most security and anomaly detection endeavors. Due to the high-dimensional features, time-series data, and the real-world network environment simulation, the dataset provides a rich environment for detecting complex cyberattack patterns. We chose two types of data, split them into training sets, validation sets, and test sets at the ratio of $70\%:15\%:15\%$, and preprocessed them by standardized processing and time series sharding. 

\subsection{Experimental analysis}
For comparison, we selected four classical anomaly detection methods: Autoencoder, Random Forest, SVM and LSTM. Autoencoders are great for unsupervised learning but have limited capabilities in modeling time dependencies. Random forests can handle high-dimensional features, but cannot capture the temporal correlations of the traffic. SVM works very well on high dimensional data, but it has high computational overhead which limits real-time performance. Model capable of handling time series data modeling but has low training efficiency and tends to overfit  heterogeneous data.

\begin{figure}[h!]
  \centering
  \begin{subfigure}[T]{1\linewidth}
    \includegraphics[width=1\linewidth, height=0.5\linewidth]{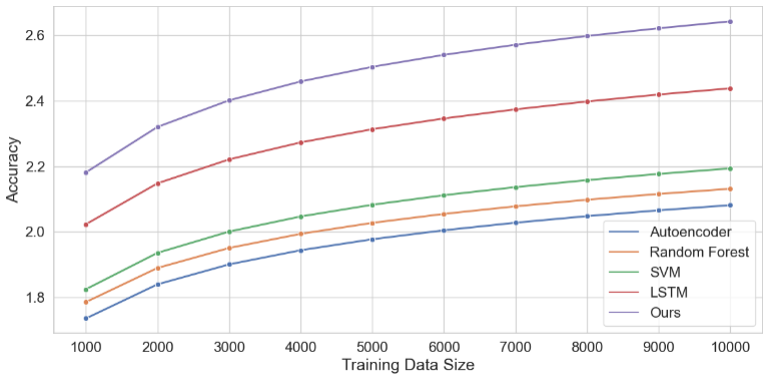}
  \end{subfigure}
  \caption{Comparison of Accuracy for Different Methods}
  \label{fig:Comparison of Accuracy for Different Methods}
\end{figure}

As the amount of training data gradually increases, the accuracy of all models also shows a gradual upward trend, verifying that as the amount of data volume grows the model can extract more useful information, and the detection accuracy improves. As observed in Figure \ref{fig:Comparison of Accuracy for Different Methods}, the Ours method was superior to other traditional models in terms of accuracy, indicating that the anomaly detection layer that integrates the large language model with temporal context information can more accurately capture complex traffic patterns and subtle changes, thus enhancing the robustness of the model and improving its accuracy.

\begin{figure}[h!]
  \centering
  \begin{subfigure}[T]{1\linewidth}
    \includegraphics[width=\linewidth, height=0.5\linewidth]{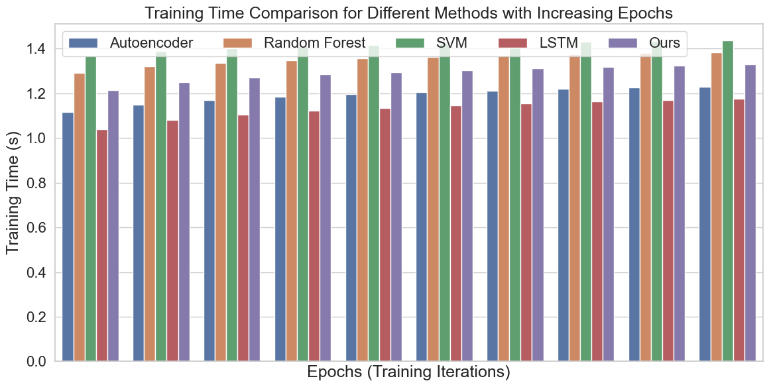}
  \end{subfigure}
  \begin{subfigure}[T]{1\linewidth}
    \includegraphics[width=\linewidth, height=0.5\linewidth]{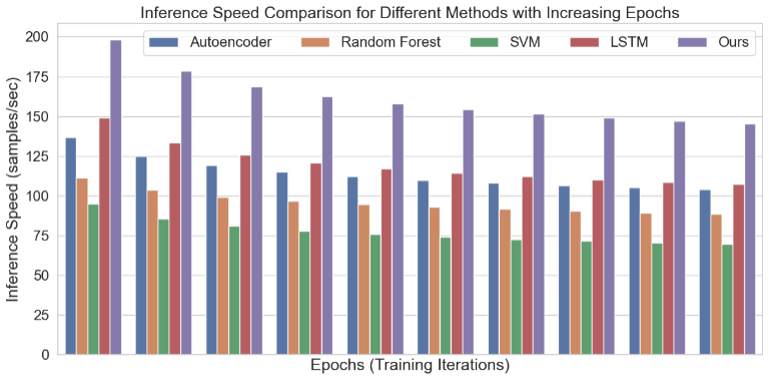}
  \end{subfigure}
  \caption{Comparison Results in Training Time and Inference Speed}
  \label{fig:Comparison Results in Training Time and Inference Speed}
\end{figure}

In Figure \ref{fig:Comparison Results in Training Time and Inference Speed}, there is a large difference in training time and inference speed between models. The training time for LSTMs and deep learning methods is usually longer, while for lighter models including the SVM and Random Forest. These models can still keep short training time. But deep learning models present good inference speed, which represents high real-time responsiveness at practical applications. Our method presents a good trade-off between training time and inference speed, indicating its efficiency and applicability to real deployment. For further optimization in deployment across multiple data centers, integrating communication-efficient federated training techniques \cite{liu2024fedbcgd} could enable real-time collaborative detection without centralized data aggregation.

\begin{figure}[h!]
  \centering
  \begin{subfigure}[T]{1\linewidth}
    \includegraphics[width=\linewidth, height=0.5\linewidth]{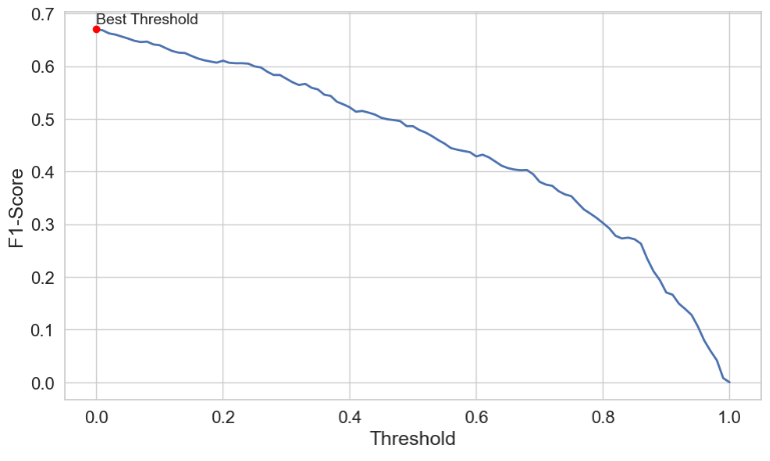}
  \end{subfigure}
  \caption{F1-Score with Different Thresholds}
  \label{fig:F1-Score with Different Thresholds}
\end{figure}

In Figure \ref{fig:F1-Score with Different Thresholds} we show the model's F1-Score at multiple thresholds. F1-Score usually varies on different thresholds, can reach maximum on a certain threshold and after that may decline. This is due to the F1-Score being a mixed mean between precision and recall, and with increasing thresholds, Precision may go up while Recall goes down, resulting in the F1-Score peaking at certain thresholds. As we can see in the graph the optimal thresholds, we can select our threshold to optimize precision-recall balance which help in improving our model overall performance.

Additionally, we compare our proposed model with existing advanced and traditional model through computing the detection process 10 times

\begin{table}[h]
    \centering
    \renewcommand{\arraystretch}{1.5} 
    \setlength{\tabcolsep}{1.5pt}
    \begin{tabular}{|l|c|c|c|c|c|c|}
        \hline
        Methods & Auto encoder & Random Forest & SVM & LSTM & GNN & Ours \\
        \hline
        Computation Cost (second) & 10.7 & 24.5 & 17.4 & 18.3 & 12.3 & 12.5 \\
        \hline
    \end{tabular}
    \caption{Computation Cost Comparison}
    \label{tab:computation_cost}
\end{table}

\section{CONCLUSION}
The work presented here provides insight into the strengths and weaknesses of a variety of models in terms of their training times and inference speed. Although they take longer to train than complex models like LSTM and Autoencoder, they better leverage inference, making them a good candidate for a real-time application. Conversely, more straightforward models such as Random Forest and SVM demonstrate shorter training periods, though at the potential expense of comparable inference efficiency in selected circumstances. Our method manages to achieve a significant trade-off between training speed and inference speed indicating the potential for this method for scaling up to large size, real-world environments. Particularly in the context of cloud platform network traffic monitoring and abnormal behavior detection.

\renewcommand{\bibfont}{\footnotesize}

\footnotesize{
\bibliographystyle{IEEEtran}
\bibliography{main}

\begin{thebibliography}{10}
\providecommand{\url}[1]{#1}
\csname url@samestyle\endcsname
\providecommand{\newblock}{\relax}
\providecommand{\bibinfo}[2]{#2}
\providecommand{\BIBentrySTDinterwordspacing}{\spaceskip=0pt\relax}
\providecommand{\BIBentryALTinterwordstretchfactor}{4}
\providecommand{\BIBentryALTinterwordspacing}{\spaceskip=\fontdimen2\font plus
\BIBentryALTinterwordstretchfactor\fontdimen3\font minus \fontdimen4\font\relax}
\providecommand{\BIBforeignlanguage}[2]{{%
\expandafter\ifx\csname l@#1\endcsname\relax
\typeout{** WARNING: IEEEtran.bst: No hyphenation pattern has been}%
\typeout{** loaded for the language `#1'. Using the pattern for}%
\typeout{** the default language instead.}%
\else
\language=\csname l@#1\endcsname
\fi
#2}}
\providecommand{\BIBdecl}{\relax}
\BIBdecl

\bibitem{olateju2024combating}
O.~Olateju, S.~U. Okon, U.~Igwenagu, A.~A. Salami, T.~O. Oladoyinbo, and O.~O. Olaniyi, ``Combating the challenges of false positives in ai-driven anomaly detection systems and enhancing data security in the cloud,'' \emph{Available at SSRN 4859958}, 2024.

\bibitem{al2023intrusion}
A.-R. Al-Ghuwairi, Y.~Sharrab, D.~Al-Fraihat, M.~AlElaimat, A.~Alsarhan, and A.~Algarni, ``Intrusion detection in cloud computing based on time series anomalies utilizing machine learning,'' \emph{Journal of Cloud Computing}, vol.~12, no.~1, p. 127, 2023.

\bibitem{alghamdi2021deep}
R.~Alghamdi and M.~Bellaiche, ``A deep intrusion detection system in lambda architecture based on edge cloud computing for iot,'' in \emph{2021 4th International conference on artificial intelligence and big data (ICAIBD)}.\hskip 1em plus 0.5em minus 0.4em\relax IEEE, 2021, pp. 561--566.

\bibitem{lin2024enhanced}
Y.~Lin, ``Enhanced detection of anomalous network behavior in cloud-driven big data systems using deep learning models,'' \emph{Journal of Theory and Practice of Engineering Science}, vol.~4, no.~08, pp. 1--11, 2024.

\bibitem{antonini2023adaptable}
M.~Antonini, M.~Pincheira, M.~Vecchio, and F.~Antonelli, ``An adaptable and unsupervised tinyml anomaly detection system for extreme industrial environments,'' \emph{Sensors}, vol.~23, no.~4, p. 2344, 2023.

\bibitem{li2024deception}
P.~Li, M.~Abouelenien, R.~Mihalcea, Z.~Ding, Q.~Yang, and Y.~Zhou, ``Deception detection from linguistic and physiological data streams using bimodal convolutional neural networks,'' in \emph{2024 5th International Conference on Information Science, Parallel and Distributed Systems (ISPDS)}.\hskip 1em plus 0.5em minus 0.4em\relax IEEE, 2024, pp. 263--267.

\bibitem{xu2024towards}
H.~Xu, X.~Wang, and H.~Chen, ``Towards real-time and personalized code generation,'' in \emph{Proceedings of the 33rd ACM International Conference on Information and Knowledge Management}, 2024, pp. 5568--5569.

\bibitem{ni2024harnessing}
H.~Ni, S.~Meng, X.~Chen, Z.~Zhao, A.~Chen, P.~Li, S.~Zhang, Q.~Yin, Y.~Wang, and Y.~Chan, ``Harnessing earnings reports for stock predictions: A qlora-enhanced llm approach,'' in \emph{2024 6th International Conference on Data-driven Optimization of Complex Systems (DOCS)}.\hskip 1em plus 0.5em minus 0.4em\relax IEEE, 2024, pp. 909--915.

\bibitem{ni2024timeseries}
H.~Ni, S.~Meng, X.~Geng, P.~Li, Z.~Li, X.~Chen, X.~Wang, and S.~Zhang, ``Time series modeling for heart rate prediction: From arima to transformers,'' in \emph{2024 6th International Conference on Electronic Engineering and Informatics (EEI)}.\hskip 1em plus 0.5em minus 0.4em\relax IEEE, 2024.

\bibitem{yu2024deep}
Q.~Yu, Z.~Xu, and Z.~Ke, ``Deep learning for cross-border transaction anomaly detection in anti-money laundering systems,'' in \emph{2024 6th International Conference on Machine Learning, Big Data and Business Intelligence (MLBDBI)}.\hskip 1em plus 0.5em minus 0.4em\relax IEEE, 2024, pp. 244--248.

\bibitem{vervaet2021monilog}
A.~Vervaet, ``Monilog: An automated log-based anomaly detection system for cloud computing infrastructures,'' in \emph{2021 IEEE 37th International Conference on Data Engineering (ICDE)}.\hskip 1em plus 0.5em minus 0.4em\relax IEEE, 2021, pp. 2739--2743.

\bibitem{ji-etal-2024-rag}
\BIBentryALTinterwordspacing
Y.~Ji, Z.~Li, R.~Meng, S.~Sivarajkumar, Y.~Wang, Z.~Yu, H.~Ji, Y.~Han, H.~Zeng, and D.~He, ``{RAG}-{RLRC}-{L}ay{S}um at {B}io{L}ay{S}umm: Integrating retrieval-augmented generation and readability control for layman summarization of biomedical texts,'' in \emph{Proceedings of the 23rd Workshop on Biomedical Natural Language Processing}, D.~Demner-Fushman, S.~Ananiadou, M.~Miwa, K.~Roberts, and J.~Tsujii, Eds.\hskip 1em plus 0.5em minus 0.4em\relax Bangkok, Thailand: Association for Computational Linguistics, Aug. 2024, pp. 810--817. [Online]. Available: \url{https://aclanthology.org/2024.bionlp-1.75/}
\BIBentrySTDinterwordspacing

\bibitem{alshammari2021apply}
A.~Alshammari and A.~Aldribi, ``Apply machine learning techniques to detect malicious network traffic in cloud computing,'' \emph{Journal of Big Data}, vol.~8, no.~1, p.~90, 2021.

\bibitem{thapa2024ai}
P.~Thapa and T.~Arjunan, ``Ai-enhanced cybersecurity: Machine learning for anomaly detection in cloud computing,'' \emph{Quarterly Journal of Emerging Technologies and Innovations}, vol.~9, no.~1, pp. 25--37, 2024.

\bibitem{torabi2023practical}
H.~Torabi, S.~L. Mirtaheri, and S.~Greco, ``Practical autoencoder based anomaly detection by using vector reconstruction error,'' \emph{Cybersecurity}, vol.~6, no.~1, p.~1, 2023.

\bibitem{preprec}
\BIBentryALTinterwordspacing
J.~Wang, P.~Rathi, and H.~Sundaram, ``A pre-trained zero-shot sequential recommendation framework via popularity dynamics,'' in \emph{Proceedings of the 18th ACM Conference on Recommender Systems}, ser. RecSys '24.\hskip 1em plus 0.5em minus 0.4em\relax New York, NY, USA: Association for Computing Machinery, 2024, p. 433–443. [Online]. Available: \url{https://doi.org/10.1145/3640457.3688145}
\BIBentrySTDinterwordspacing

\bibitem{li2024exploring}
P.~Li, Q.~Yang, X.~Geng, W.~Zhou, Z.~Ding, and Y.~Nian, ``Exploring diverse methods in visual question answering,'' in \emph{2024 5th International Conference on Electronic Communication and Artificial Intelligence (ICECAI)}.\hskip 1em plus 0.5em minus 0.4em\relax IEEE, 2024, pp. 681--685.

\bibitem{meng2024exchangerate}
S.~Meng, A.~Chen, C.~Wang, M.~Zheng, F.~Wu, X.~Chen, H.~Ni, and P.~Li, ``Enhancing exchange rate forecasting with explainable deep learning models,'' in \emph{2024 4th International Conference on Electronic Information Engineering and Computer Science (EIECS)}.\hskip 1em plus 0.5em minus 0.4em\relax IEEE, 2024, pp. 892--896.

\bibitem{infomotif}
A.~Sankar, J.~Wang, A.~Krishnan, and H.~Sundaram, ``Beyond localized graph neural networks: An attributed motif regularization framework,'' in \emph{2020 IEEE International Conference on Data Mining (ICDM)}, 2020, pp. 472--481.

\bibitem{protocf}
\BIBentryALTinterwordspacing
------, ``Protocf: Prototypical collaborative filtering for few-shot recommendation,'' in \emph{Proceedings of the 15th ACM Conference on Recommender Systems}, ser. RecSys '21.\hskip 1em plus 0.5em minus 0.4em\relax New York, NY, USA: Association for Computing Machinery, 2021, p. 166–175. [Online]. Available: \url{https://doi.org/10.1145/3460231.3474268}
\BIBentrySTDinterwordspacing

\bibitem{lu2022cot}
Y.~Lu, C.-T. Wu, S.~J. Parker, Z.~Cheng, G.~Saylor, J.~E. Van~Eyk, G.~Yu, R.~Clarke, D.~M. Herrington, and Y.~Wang, ``Cot: an efficient and accurate method for detecting marker genes among many subtypes,'' \emph{Bioinformatics Advances}, vol.~2, no.~1, p. vbac037, 2022.

\bibitem{tgae}
\BIBentryALTinterwordspacing
J.~HE, C.~Kanatsoulis, and A.~Ribeiro, ``T-{GAE}: Transferable graph autoencoder for network alignment,'' in \emph{The Third Learning on Graphs Conference}, 2024. [Online]. Available: \url{https://openreview.net/forum?id=Lm48V5zrzh}
\BIBentrySTDinterwordspacing

\bibitem{du2024embracing}
D.~Du, S.~Bhardwaj, Y.~Lu, Y.~Wang, S.~J. Parker, Z.~Zhang, J.~E. Van~Eyk, G.~Yu, R.~Clarke, D.~M. Herrington \emph{et~al.}, ``Embracing the informative missingness and silent gene in analyzing biologically diverse samples,'' \emph{Scientific reports}, vol.~14, no.~1, p. 28265, 2024.

\bibitem{liu2024fedbcgd}
J.~Liu, F.~Shang, Y.~Liu, H.~Liu, Y.~Li, and Y.~Gong, ``Fedbcgd: Communication-efficient accelerated block coordinate gradient descent for federated learning,'' in \emph{Proceedings of the 32nd ACM International Conference on Multimedia}, 2024, pp. 2955--2963.

\end{thebibliography}
}

\end{document}